\documentclass[showpacs,aps,twocolumn,nofootinbib,11]{revtex4}
\usepackage{amsmath}
\usepackage{epsfig}
\usepackage{subfigure}
\usepackage{float}
\usepackage{longtable}
\usepackage{times}
\usepackage{txfonts}
\usepackage{color}
\usepackage{ulem}

 \usepackage{ifpdf}
 \ifpdf
 \usepackage[pdftex]{hyperref}
 \else
 \usepackage[hypertex]{hyperref}
 \fi

 \hypersetup{
   pdftitle={},%
   pdfauthor={},%
   pdfsubject={},%
   pdfkeywords={},%
   pdfstartview={},%
   bookmarksopen=true, breaklinks=true, debug=true, %
   colorlinks=true, linkcolor=red, citecolor=blue, urlcolor=blue
 }

\newcommand{\red}[1]{\textcolor[rgb]{0,0,0}{#1}}
\newcommand{\green}[1]{\textcolor[rgb]{0,0,0}{#1}}

\newcommand{\beq}{\begin{eqnarray}}
\newcommand{\eeq}{\end{eqnarray}}
\newcommand{\be}{\begin{eqnarray*}}
\newcommand{\ee}{\end{eqnarray*}}

\def\lsim{\raise0.3ex\hbox{$<$\kern-0.75em\raise-1.1ex\hbox{$\sim$}}}
\def\gsim{\raise0.3ex\hbox{$>$\kern-0.75em\raise-1.1ex\hbox{$\sim$}}}

\def\d+Au  {$d$Au}
\def\d+Aum  {d\mathrm{Au}}

\def\beq     {\begin{equation}}
\def\eeq     {\end{equation}}

\long\def\symbolfootnote[#1]#2{\begingroup%
\def\thefootnote{\fnsymbol{footnote}}\footnote[#1]{#2}\endgroup}

\begin{document}


\title{Excited charmonium suppression in proton-nucleus collisions as a consequence of 
comovers}

\author{E. G.~Ferreiro}

\affiliation{Departamento de F{\'\i}sica de Part{\'\i}culas, Universidade de Santiago de Compostela, 15782 Santiago de Compostela, Spain}

\begin{abstract}
Recent results from proton(deuteron)-nucleus collisions at RHIC and LHC energies 
have shown an 
unexpected suppression of excited
quarkonium states as compared to their ground states.
In particular,
stronger
suppression of the  $\psi(2S)$ relative to the $J/\psi$ 
has been detected.
Similar observations were made at lower
energies and were easily explained by nuclear absorption. At higher energies, a similar
explanation would violate the Heisenberg principle, \red{since the calculations based on the uncertainty principle lead to a charmonium formation
time expected to be larger than the nuclear radius, which results in identical nuclear break-up probability
for the $\psi(2S)$ and $J/\psi$. }
On the contrary, this behavior is naturally explained by the interactions of the quarkonium states with a comoving medium.
We present our results on $J/\psi$ and $\psi(2S)$ production for d+Au collisions at $\sqrt{s}=200$ GeV and for p+Pb collisions at $\sqrt{s}=5.02$ TeV. 
\end{abstract}

\pacs{13.85.Ni,14.40.Pq,21.65.Jk,25.75.Dw}
\maketitle

\red{Charmonium} mesons have captured our attention for decades. 
Due to the high scale provided by their large masses, 
they are considered to be outstanding 
probes of Quantum Chromodynamics (QCD).
The interest in this field concerns the issue of their production mechanisms in proton-proton collisions 
together with their interaction with the nuclear matter created in ultrarelativistic heavy-ion collisions.
\vskip 0.15cm

This is so since 
lattice QCD calculations predict that, at
sufficiently large energy densities, hadronic matter undergoes a
phase transition to a plasma of deconfined quarks and
gluons, the so-called quark gluon plasma (QGP), where
the QCD binding potential is screened. 
Given the existence
of several quarkonium states, each of them with different
binding energies, they are expected to sequentially
melt into open charm or bottom mesons above certain
energy density thresholds. Thus,
the production and absorption of quarkonium in a nuclear
medium provide quantitative inputs for the study
of QCD at high density and temperature. 
\vskip 0.15cm

The interest on quarkonium production is not restricted to the study of deconfinement.
Puzzling features in proton(deuteron)-nucleus collision data, where the deconfinement cannot be reached, reveal new
aspects of charmonium physics in nuclear reactions,
namely the role of cold nuclear matter effects. 
\vskip 0.15cm

In particular, the measurement
of production rates for multiple quarkonium states, with
different physical sizes and binding energies, offers an excellent tool for probing
the time scale of the evolution of heavy quark-antiquark
pairs into bound color singlet quarkonium states which represents
a challenge within QCD.
\vskip 0.15cm

As a matter of fact, recent unexpected results on $\psi(2S)$ production in d+Au and p+Pb collisions from PHENIX \cite{Adare:2013ezl} and ALICE  \cite{Abelev:2014zpa,Arnaldi:2014kta} Collaborations have shown an important suppression of its yield with respect to proton-proton production. Furthermore, this suppression is stronger than the one previously obtained for the $J/\psi$ production. 
Former measurements of $J/\psi$ and $\psi(2S)$ production rates in proton-nucleus collisions at lower energies by E866/NuSea \cite{Leitch:1999ea} and by NA50 \cite{Alessandro:2006jt} Collaborations also show a stronger suppression of the excited state near $x_F \approx 0$. At those lower energies, this dissimilarity has been interpreted as the effect of $c\bar{c}$ break-up in interactions with \red{the primordial} nucleons, the so-called nuclear absorption. 
When the time spent
traversing the nucleus by the $c\bar{c}$ pair becomes longer than
the charmonia formation time, the larger $\psi(2S)$ meson will
be further suppressed by a stronger nuclear break-up effect.
\vskip 0.15cm

However, at higher energies, the charmonium formation time is expected to be larger than the nucleus radius  \cite{Ferreiro:2012mm}.
Following
the uncertainty principle,
the formation time is related
to the time needed --in their rest frame-- to distinguish the energy levels of the
$1S$ and $2S$ states,
$\red{\tau}_f=  \frac{2 M_{c\bar c}}{(M^2_{2S}-M^2_{1S})}= 2 \times 3.3$ GeV / 4 GeV$^2= 0.35$ fm
for the $\psi$.
Moreover, \red{this formation time} 
has to be considered in the rest frame of the target nucleus,
{\it i.e.}~the Au beam at RHIC and the Pb beam at LHC.
\red{In this case, the formation time is increased by the Lorentz boost factor, $t_f=\gamma\ \tau$.}
The boost factor $\gamma$ is obtained from the rapidity of the pair corrected by the \red{nucleus} beam rapidity,
$\gamma=\cosh(y-y_{beam}^{\rm Au})$ where $y_{beam}^{\rm Au}=\red{-ln(\frac{\sqrt{s}}{m_N})}=-5.36$ for RHIC,
\red{resulting in a boost factor bigger than 100 at mid and forward rapidities \cite{Ferreiro:2012mm}.}
\red{Consequently}, in the mid and forward rapidity regions at RHIC, $t_f$ is significantly larger than
the Au radius, $t_f=36.7$ fm for $y=0$ and larger for forward rapidities. 
At the LHC, for a lead beam of 1.5\red{8} TeV in pPb mode\footnote{\red{Due to the LHC design, the colliding beams have different energies per nucleon, $E_{\rm p}=4$ TeV, $E_{\rm Pb}=1.58$ TeV,  \green{and cannot
be tuned separately.} As a consequence, the centre of
mass of the nucleon-nucleon collision is shifted by $\Delta y=0.465$ with respect to the laboratory frame in the
direction of the proton beam \cite{Abelev:2014zpa}.}}, $y_{beam}^{\rm \red{Pb}}=-8.11$, which 
\red{results in a boost of more than 1000 at LHC mid rapidities \cite{Ferreiro:2013pua}.} This implies that the chamonium formation time will be larger 
than the Pb radius practically in the whole rapidity region.
This means that the $c {\bar c}$ is nearly always in a
pre-resonant state when traversing the nuclear matter, which results in identical break-up probability 
for the $\psi(2S)$ and $J/\psi$, since these
states cannot be distinguished at the time they traverse the nucleus\footnote{Moreover, this nuclear absorption can be taken as negligible at the LHC energies.}.
\vskip 0.15cm

Other usual explanations, as the one based on
the shadowing of the heavy pairs due to the  modification of the gluon parton distribution functions in the nucleus, cannot be invoked here, since the nuclear parton shadowing effects are indistinguishable between $\psi(2S)$ and $J/\psi$ \cite{Ferreiro:2012mm}.
\vskip 0.15cm

Here, we will demonstrate
that the final state interactions of the $c\bar{c}$ pair 
with the dense medium created in the collision can cause the puzzling anomalies seen
in quarkonium production, i.e. the stronger $\psi(2S)$ suppression relative to the $J/\psi$,  within the so-called comover scenario.
In a comover framework, the suppression arises from scattering of the nascent $\psi$
with produced particles --the comovers-- 
that happen to travel along with the $c\bar{c}$ pair \cite{Gavin:1996yd,Capella97}.
\red{By comover interaction one means the interaction of the $J/\psi$ and $\psi(2S)$ particle with the produced medium: the quarkonium particle is comoving with the soft particles produced in the collision, 
their formation times being both boosted by Lorentz dilation.
This implies that the comovers can continue to interact well outside
of the nuclear volume, playing an important role. }

\vskip 0.15cm

Let us \red{recall} two common features of the comover approaches.
First,
the comover dissociation affects strongly the $\psi(2S)$ relative to the $J/\psi$, due to the larger size of the first.
Second,
the comover suppression is stronger where the comover densities are larger, i.e. it
increases with centrality and, for asymmetric collisions as proton(deuteron)-nucleus, it will be stronger in the nucleus-going direction.
\vskip 0.15cm

In the following, we will show that taking into account the above features we obtain a surprisingly good and coherent quantitative description of the available deuteron-nucleus and proton-nucleus data at RHIC and LHC energies.
We will apply the well established comover interaction model (CIM) \cite{Capella97,Armesto98,Armesto99,Capella00,Capella05,Capella:2006mb}. 
In this model, the rate equation that governs the density of charmonium at a given 
transverse coordinate $s$, impact parameter $b$ and rapidity~$y$ , $\rho^{\psi}(b,s,y)$, obeys the simple expression
\beq
\label{eq:comovrateeq}
\tau \frac{\mbox{d} \rho^{\psi}}{\mbox{d} \tau} \, \left( b,s,y \right)
\;=\; -\sigma^{co-\psi}\; \rho^{co}(b,s,y)\; \rho^{\psi}(b,s,y) \;,
\eeq
where $\sigma^{co-\psi}$ is the cross section of charmonium dissociation
due to interactions with the comoving medium of transverse density~$\rho^{co}(b,s,y)$.
\vskip 0.15cm

In order to obtain the survival probability $S^{co}_{\psi}(b,s,y)$ of a $\psi$ interacting with comovers, 
this equation is to be integrated between initial time $\tau_0$ and freeze-out time $\tau_f$.
We consider longitudinal boost invariance and neglect transverse expansion 
\red{since we have estimated that 
the transverse expansion, unlike the longitudinal one, is a very smooth process that takes place later\footnote{\red{The effect of a small transverse expansion can presumably be taken into account
by a small change of the final state interaction cross sections.}}.}
\red{Thus,} assuming a dilution
in time of the densities due to longitudinal motion which leads to a $\tau^{-1}$ dependence on
proper time, the equation can be solved analytically. The result depends only on the ratio $\tau_f/ \tau_0$
of final over initial time. Using the inverse
proportionality between proper time and densities, we put  $\tau_f/ \tau_0= \rho^{co}(b, s, y)/\rho_{pp}(y)$, i.e. we assume that the interaction stops when the densities
have diluted, reaching the value of the p+p density at the same energy. Thus, the solution of eq.~(\ref{eq:comovrateeq}) is given by
\beq
\label{eq:survivalco}
S^{co}_{\psi}(b,s,y)  \;=\; \exp \left\{-\sigma^{co-\psi}
  \, \rho^{co}(b,s,y)\, \ln
\left[\frac{\rho^{co}(b,s,y)}{\rho_{pp} (y)}\right] \right\} \;,
\eeq
where the argument of the log is the interaction time of the $\psi$ with the comovers.
\vskip 0.15cm

The main ingredient in order to
compute the survival probability $S^{co}_{\psi}$ of the quarkonium due to interactions with the comoving medium is
the density of comovers  $\rho^{co}$.
This density is not a free parameter, since it has the constraint that the total rapidity distribution $dN/dy$
of the observed particles must be reproduced.
We take it as proportional to the number of collisions,
\beq
\label{eq:ncom}
\rho^{co} (b,s,y)~=~n(b,s)\, S^{sh}_{co} (b,s)\, \frac{3}{2} (dN^{pp}_{ch}/{dy})\;,
\eeq
where $n(b,s)$ corresponds to
the number of
binary nucleon-nucleon collisions 
per unit transverse area 
at a given impact parameter, $S^{sh}_{co}$ refers to the shadowing of the parton distribution functions in a nucleus that affects the comover multiplicity,
$ch$ refers to the charged particle density in p+p and the factor $3/2$ takes into account the neutral comovers.
In order to compute the comover densities in nuclear collisions we have introduced the shadowing corrections that affects the comover multiplicities \cite{Capella99,Armesto03,Capella:2011vi}. Within this approach, a good description of the centrality dependence of charged multiplicities in nuclear collisions is obtained both
at RHIC \cite{Capella:2000dn} and LHC energies \cite{Capella:2011vi}.
\vskip 0.15cm

Finally, the comover density in p+p collisions is given by $\rho_{pp}(y)=\frac{3}{2} (dN^{pp}_{ch}/{dy})/ \pi R_p^2$, where $R_p$ is the proton radius.
We apply the experimental values and theoretical extrapolations \cite{Capella:2011vi} for the charged particle multiplicities in proton-proton collisions.
We get, at mid rapidity, the values $\rho_{pp}(0)=~2.24$~fm$^{-2}$ at $\sqrt{s}=200$~GeV \cite{Capella05} and $\rho_{pp}(0)=~3.37$~fm$^{-2}$ at $\sqrt{s}=5.02$~TeV, which correspond to the values of charged particle multiplicities $\mbox{d}N^{ch}_{pp}/\mbox{d}\eta=2$ at $\sqrt{s}=200$~GeV and $\mbox{d}N^{ch}_{pp}/\mbox{d}\eta=4.5$ at $\sqrt{s}=5.02$~TeV.
\vskip 0.15cm

\red{With the numbers quoted above, one obtains for the comover multiplicities the values 
$\mbox{d}N^{co}_{dAu}/\mbox{d}\eta=15.75$ for minimum bias d+Au collisions at $\sqrt{s}=200$ GeV at mid rapidity,
and $\mbox{d}N^{co}_{pPb}/\mbox{d}\eta=26.4$ at mid rapidity, $\mbox{d}N^{co}_{pPb}/\mbox{d}\eta=22.5$ in the p-going direction, $2.03<y<3.53$, and $\mbox{d}N^{co}_{pPb}/\mbox{d}\eta=31.2$  in the Pb-going direction, $-4.46<y<-2.96$, for minimum bias p+Pb collisions at $\sqrt{s}=5.02$ TeV. These values are consistent with the experimental charged particle multiplicities \cite{Back:2003hx,ALICE:2012xs}.}
\vskip 0.15cm

The only adjustable parameter of the comover interaction model is the cross section of charmonium dissociation due to interactions with the comoving medium, $\sigma^{co-\psi}$. It was fixed \cite{Armesto98} from fits to low-energy experimental data to be
$\sigma^{co-J/\psi}=0.65$ mb for the $J/\psi$ and $\sigma^{co-\psi(2S)}=6$ mb for the $\psi(2S)$. 
\red{The magnitude of the charmonium absorption cross section in medium is
theoretically not well under control. Different theoretical calculations of the $J/\psi$-hadron cross section based on the multipole expansion in QCD \cite{Bhanot:1979vb}
differ from those which include other non-perturbative effects by orders of magnitude \cite{Martins:1994hd}.
Moreover, its energy behavior can be quite different. 
More recent theoretical calculations based on QCD sum rules \cite{Duraes:2002ux} or on chiral effective theory using a chiral quark model \cite{Bourque:2008es}
show a moderate increase of the cross section with the energy above threshold.}
\red{On the other hand,} an invariant dissociation cross section of charmonium
on light mesons, assumed to be energy independent, is a common feature to various comover models \cite{Vogt:1999cu}.
\red{We are aware that it could change when the energy increases. We do not expect this effect to be extremely important above $\sqrt{s}=20$ GeV --well above threshold-- and, since we are
unable to evaluate the magnitude of this eventual change,} 
we have chosen to keep it fixed to its low energy value.
This value has been also successfully applied at higher energies to reproduced the RHIC \cite{Capella:2007jv} and LHC \cite{Ferreiro:2012rq} data on $J/\psi$ from nucleus-nucleus collisions.
\vskip 0.15cm

Note that, together with the comover interaction, another important effect that plays a role in quarkonium nuclear production is the initial shadowing of the heavy pairs due to the  modification of the gluon parton distribution functions in the nucleus. It can be calculated analytically in the above mentioned framework or using any of the available parametrizations for the nuclear parton distribution functions  \cite{Ferreiro:2013pua,Ferreiro:2012sy}. Note also that this effect is to be taken identical for the states 1S and 2S \cite{Ferreiro:2012mm}, i.e. for the $J/\psi$ and the $\psi(2S)$.
This nuclear modification of the parton distribution functions will produce a common decrease of the $J/\psi$ and the $\psi(2S)$ yields in the mid and forward rapidity regions both at RHIC and LHC energies. It can induce an increase of both yields in the backward rapidity region.
\vskip 0.15cm

It is now straightforward to calculate the nuclear modification factor, 
i.e. the ratio of the $\psi$ yield in proton(deuteron)-nucleus collisions to the $\psi$ yield in proton-proton collisions
multiplied by the average number of binary nucleon-nucleon collisions:
\begin{multline}
\label{eq:ratiopsi}
R^{\psi}_{pA}(b) \;=\;
\frac{\mbox{d}N^{\psi}_{pA}/\mbox{d}y}{n(b)
  \,\mbox{d}N^{\psi}_{pp}/\mbox{d}y} \\ 
\;=\; \frac{\int\mbox{d}^2s \, 
  \sigma_{pA}(b) \, n(b,s) \,  S_{\psi}^{sh}(b,s) \, S^{co}_{\psi}(b,s)
}{\int \mbox{d}^2 s \, \sigma_{pA} (b) \, n(b,s)} \;, 
\end{multline}
where $S^{co}_{\psi}$ refers to the survival probability due to the medium interactions while $S_{\psi}^{sh}$ takes into account 
the shadowing of the parton distribution functions in a nucleus that affects the $\psi$ production.
Any nuclear effect affecting quarkonium production leads to a deviation
of $R_{pA}$ from unity.
\vskip 0.15cm

\begin{figure}[thb]
\includegraphics[width=1.0\linewidth]{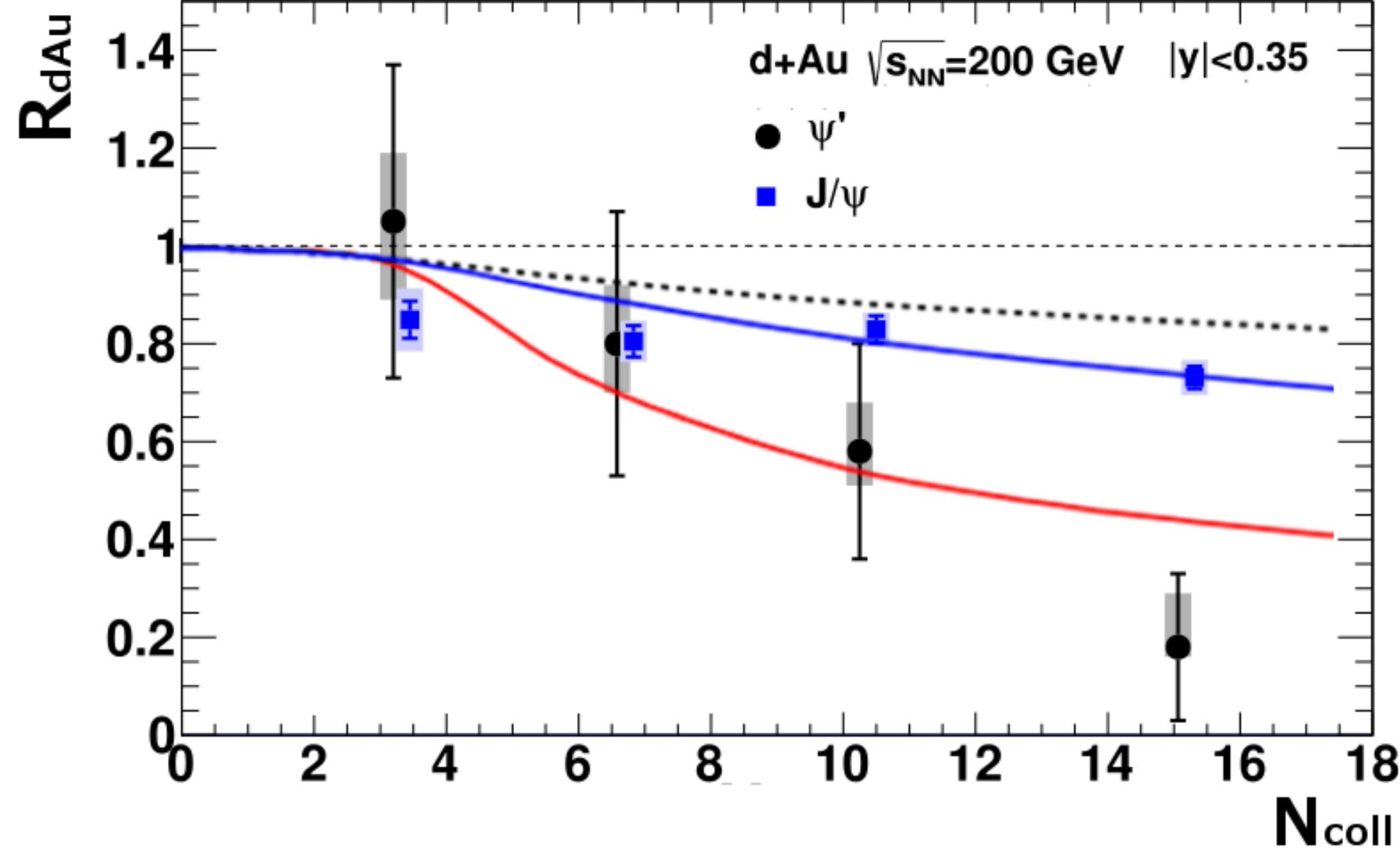}
\caption{\label{fig:figdAu}(Color online) 
The $J/\psi$ (blue continuous line) and $\psi(2S)$ (red continuous line)
nuclear modification factor $R_{dAu}$ as a function of the number of collisions $N_{coll}$
compared to the PHENIX data \cite{Adare:2013ezl}.  The suppression due to the shadowing corrections (discontinuous line) is also shown. 
}
\end{figure}
Figure~\ref{fig:figdAu} shows the nuclear modification factor $R_{dAu}$ as a function of the number of collisions $N_{coll}$ for the $J/\psi$ and $\psi(2S)$ production in d+Au collisions at $\sqrt{s}=200$ GeV compared to PHENIX experimental data \cite{Adare:2013ezl}. We observe a strong suppression of $\psi(2S)$ production with increasing centrality. This suppression is a factor of 2 times larger than the observed suppression for $J/\psi$ production for the most central events. The suppression due to the shadowing corrections, calculated within the EPS09 LO parametrization \cite{Eskola:2009uj} 
and identical for the $J/\psi$ and $\psi(2S)$, has been taken into account. It represents more than 50 \% of the total $J/\psi$ suppression, while the $\psi(2S)$ is mostly suppressed due to the comover interaction.
\vskip 0.15cm 

\red{Note that, 
in order to include the shadowing, the original analysis EPS09 LO has been used \cite{Eskola:2009uj}. The centrality dependence of shadowing is not addressed in this model.
It can be parameterized \cite{Vogt:2004dh,Ferreiro:2008qj} assuming that the inhomogeneous shadowing is proportional to the local density $\rho_A$ or the thickness function $T_A$,
\begin{equation}
\label{eq5}
R^A_i (b,x,Q^2)=1+[R^A_i (x,Q^2)-1]\ N_{\rho}
\frac{\int dz\ \rho_A(b,z)}{\int dz\ \rho_A(0,z)}\ ,
\end{equation}
where $b$ and $z$ are the transverse and longitudinal location in position space, $\rho_A(b,z)$ corresponds
to the Woods-Saxon distribution for the nucleon density in the nucleus, related to the nuclear profile function $T_A(b)$ by 
$\int dz\ \rho_A(b,z)= A\ T_A(b)$, $\rho_0$ is the central density, given by the normalization $\int d^2b \int dz\ \rho_A(b,z) = A$
and $R^A_i (x,Q^2)$ is the shadowing function from EPS09 LO. 
$N_{\rho}$ is the normalization 
factor, and it is chosen so that $(1/A) \int d^2b \int dz\ \rho_A(b,z)\ R^A_i (b,x,Q^2) = R^A_i (x,Q^2)$\footnote{\red{
This scenario corresponds to what is referred to as "1-parameter approach" in EPS09s \cite{Helenius:2012wd}, i.e. EPS09 with an 
spatial (impact parameter) dependence introduced in terms of powers of the nuclear thickness functions $T_A(s)$.
The 1-parameter approach (n = 1 in the power series) agrees well with the full EPS09s in the LO approximation, while it may introduces differences in the slope for the NLO approximation.
}}.}
\vskip 0.15cm

We proceed now with the analysis of $J/\psi$ and $\psi(2S)$ production in p+Pb collisions at $\sqrt{s}=5.02$ TeV, that offers an excellent opportunity to verify the role of comovers on quarkonium suppression. According to available experimental data \cite{Abelev:2014zpa,Arnaldi:2014kta}, 
the suppression of the $J/\psi$ shows a strong difference between the forward and backward rapidity regions, while the $\psi(2S)$ shows astonishing similar suppression in both rapidity intervals. 
\vskip 0.15cm
\begin{figure}[thb]
\includegraphics[width=1.\linewidth]{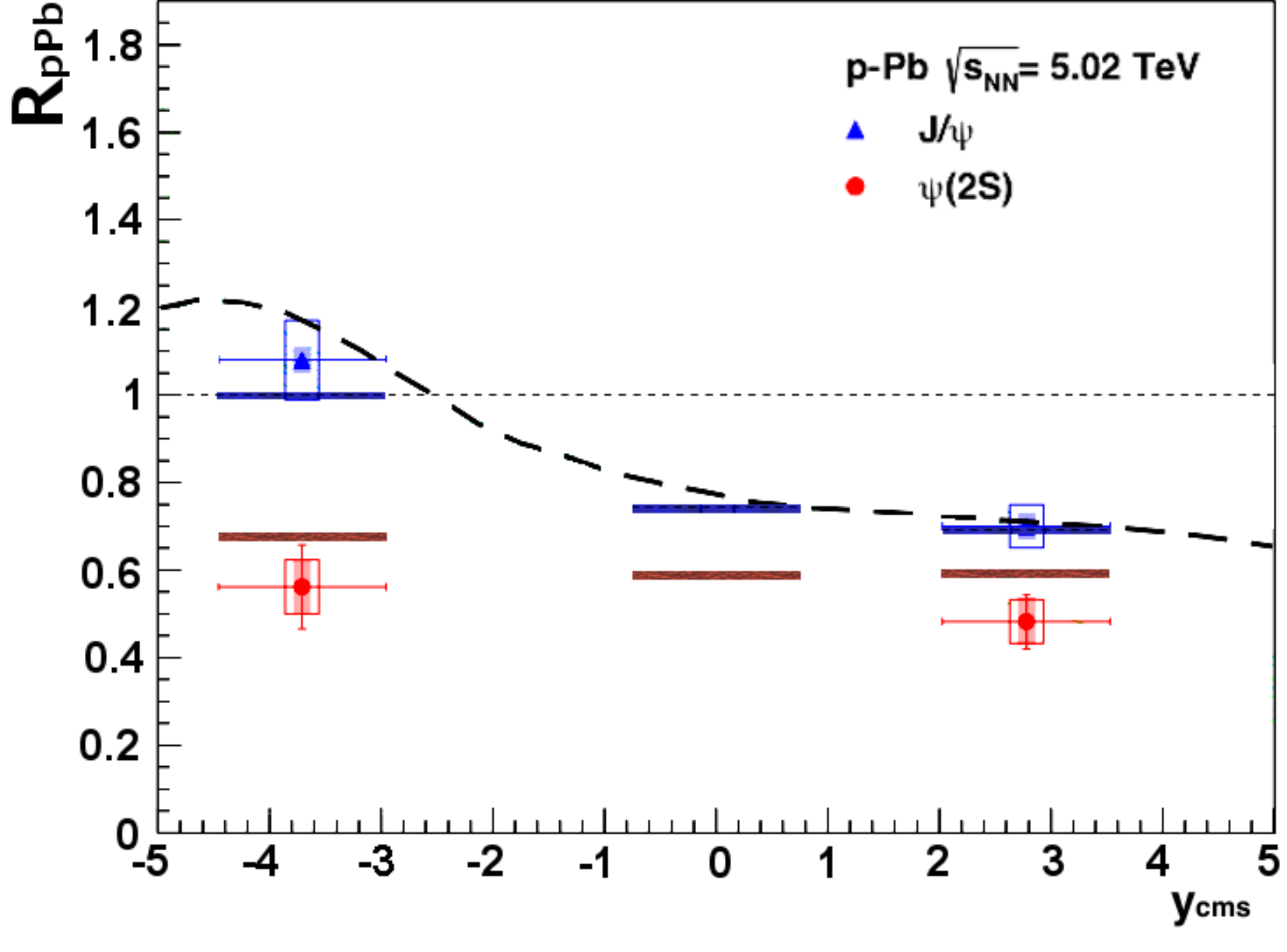}
\caption{\label{fig:figpPby}(Color online) 
The $J/\psi$ (blue line) and $\psi(2S)$ (red line)
nuclear modification factor $R_{pPb}$ as a function of rapidity
compared to the ALICE data \cite{Abelev:2014zpa}. The suppression due to the shadowing corrections (discontinuous line) is also shown.
}
\end{figure}

\begin{figure}[hb]
\vskip 0.5cm
\includegraphics[width=1\linewidth]{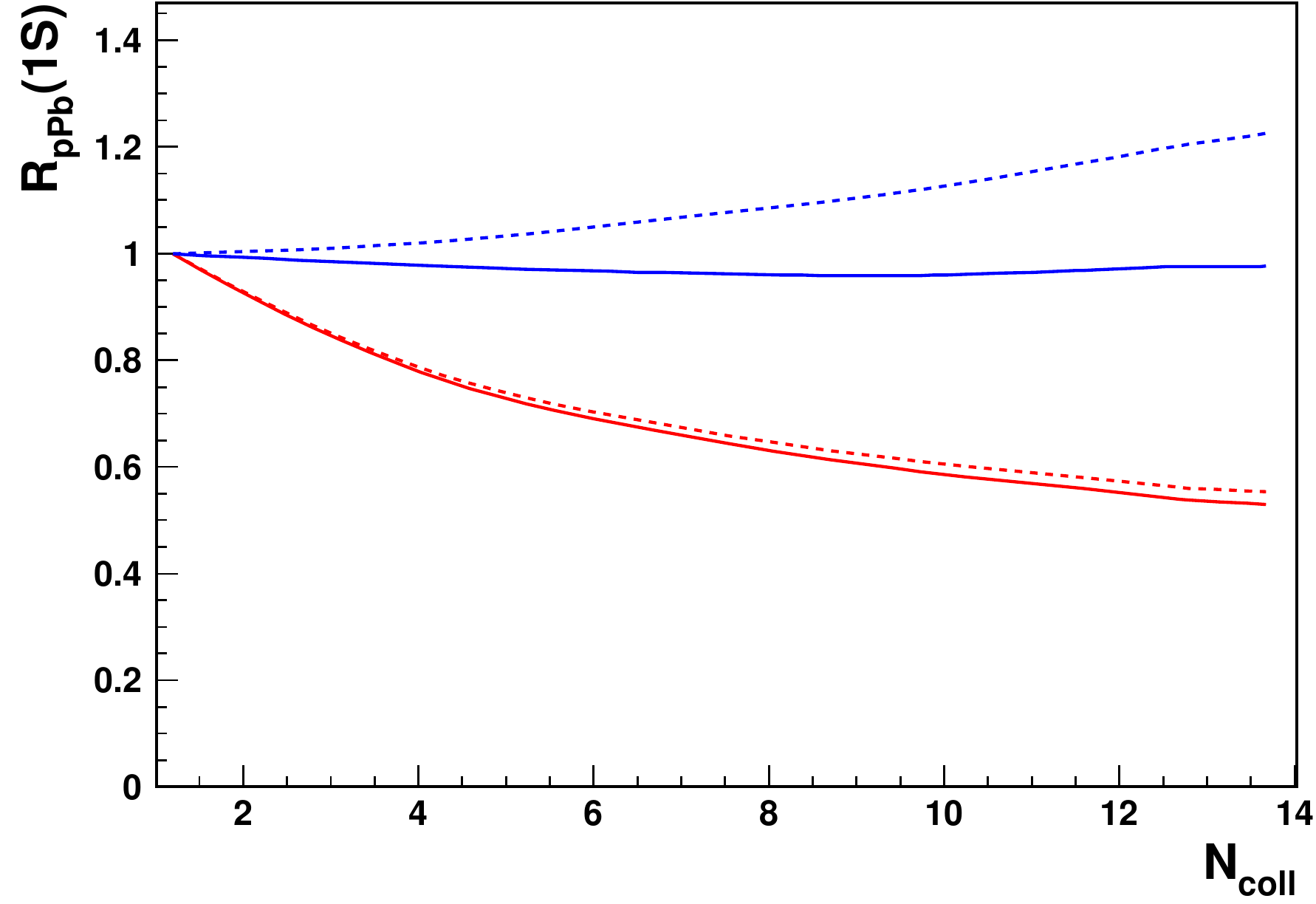}
\includegraphics[width=1\linewidth]{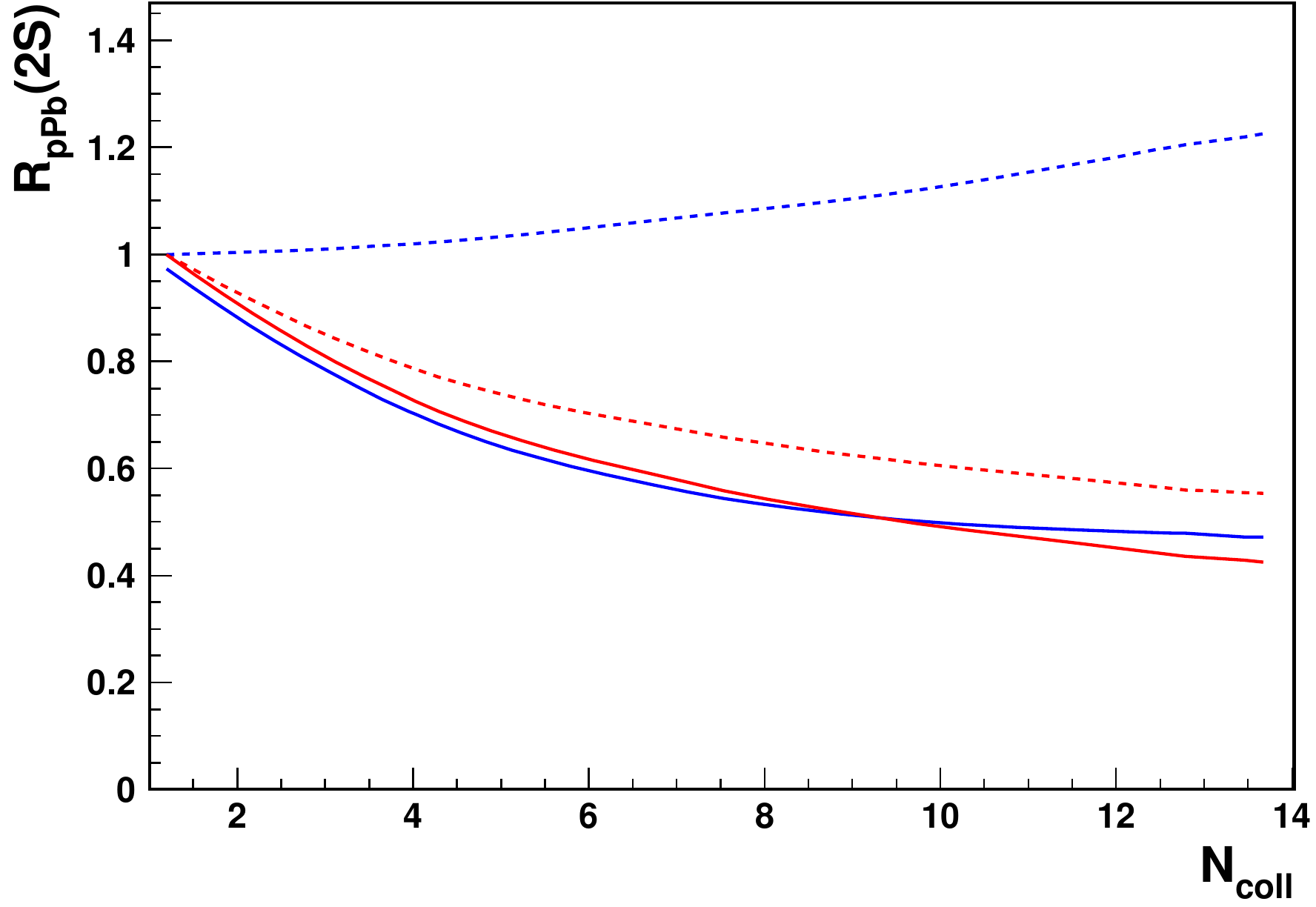}
\caption{\label{fig:figpPbvncoll}(Color online) 
The $J/\psi$ (upper figure) and $\psi(2S)$ (lower figure)
nuclear modification factor $R_{pPb}$ as a function of the number of collisions
in the backward $-4.46<y<-2.96$ (blue continuous line) and forward $2.03<y<3.53$ (red continuous line) rapidity intervals. 
The modification due to the antishadowing corrections in the backward region (blue discontinuous line) and to the shadowing corrections in the forward region (red discontinuous line) is also shown.
}
\end{figure}Note that any effect related to nuclear shadowing would produce an slight antishadowing in the backward region while it would induce a suppression in the forward region, these effects being identical for the $J/\psi$ and $\psi(2S)$.
\vskip 0.15cm

In fact, the experimental finding of a different suppression for the $\psi(2S)$ relative to the $J/\psi$, in particular in the backward rapidity region, is a clear indication of comover interactions. Actually, the density of comovers is smaller in the forward region --p-going direction-- than in the backward region --Pb-going direction--, the difference increasing with centrality, which is easily confirmed by experimental data on charged particle multiplicities  \cite{ALICE:2012xs}.
As a consequence, the effect of comovers --which differs on the $J/\psi$ and $\psi(2S)$ case-- will be strong in the backward region, while the suppression found in the forward region will be mainly due to the initial shadowing of the nuclear parton distribution functions, identical in both cases. Thus, one should expect \red{more} similar $J/\psi$ and $\psi(2S)$ suppression in the forward than in the backward region.
\vskip 0.15cm

This is illustrated in Figure~\ref{fig:figpPby}, where experimental data \cite{Abelev:2014zpa} on $J/\psi$ and $\psi(2S)$ production in p+Pb collisions at $\sqrt{s}=~5.02$~TeV are compared to our results. We have considered a common EPS09 LO shadowing \cite{Ferreiro:2013pua,Eskola:2009uj} for both the $J/\psi$ and $\psi(2S)$. The interaction with comovers, mostly at play in the backward region, is able to explained the stronger $\psi(2S)$ suppression.
\vskip 0.15cm

Our results for $J/\psi$ and $\psi(2S)$ production versus centrality in p+Pb collisions at $\sqrt{s}=5.02$ TeV are shown in Figure~\ref{fig:figpPbvncoll}.
Two rapidity intervals are studied, the p-going direction, $2.03<y<3.53$ and the Pb-going direction, $-4.46<y<-2.96$.
The effect of the EPS09 LO shadowing is completely different depending on the rapidity interval considered. While it induces an increase --antishadowing-- in the backward region, it produces a suppression --shadowing-- in the forward region.
On the other hand, the interaction with comovers introduces a suppression, stronger in backward region due to the higher comover density.
Their effect will be more important on the $\psi(2S)$ than on the $J/\psi$ production, due to the higher $\sigma^{co-\psi}$ of the first.
In fact, we obtain a nuclear modification factor $R_{pPb}^{J/\psi}$ compatible with one for the $J/\psi$ in the backward region resulting for
the combined effect of EPS09 LO nuclear modification together with the comover suppression, while the total suppression of the $J/\psi$ in the forward region
achieves almost 50\%, mainly due to the shadowing effect.
\vskip 0.15cm

Concerning the $\psi(2S)$ production, we obtain a similar suppression for the backward and forward rapidity regions. Note, nevertheless, that the origin of this decrease corresponds to different effects depending on the region of consideration: in the backward region, there is an antishadowing identical to one previously found for the $J/\psi$ which is hidden by the strong effect of comover suppression in this region; on the other side, in the forward region, both the shadowing and a limited comover effect contribute to the suppression.
\vskip 0.15cm

\begin{figure}[thb]
\includegraphics[width=1.03\linewidth]{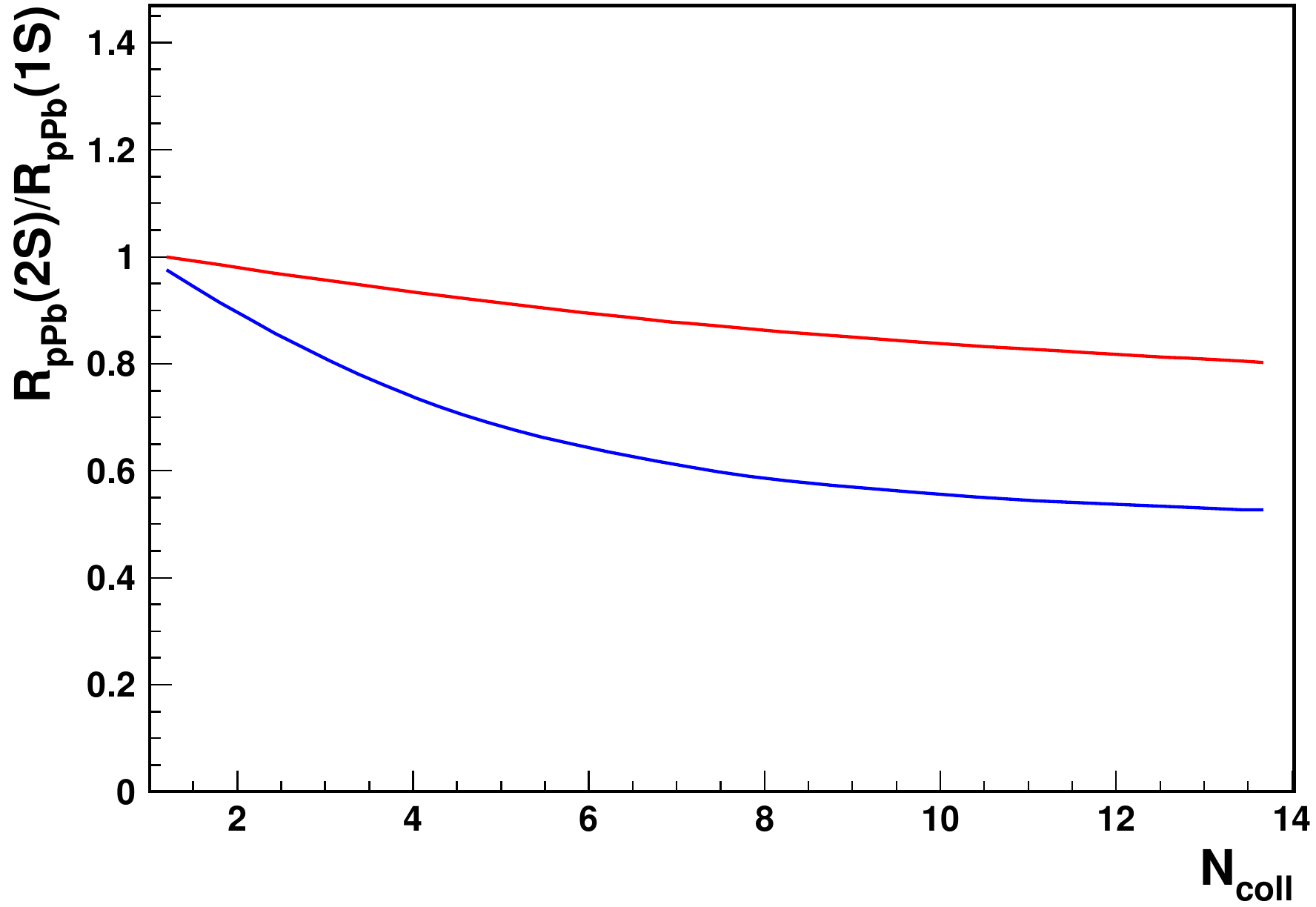}
\caption{\label{fig:figratiopPbvncoll}(Color online) 
The ratio of the $\psi(2S)$ over $J/\psi$ nuclear modification factors $R_{pPb}^{\psi(2S)}/R_{pPb}^{J/\psi}$
as a function of the number of collisions
in the backward $-4.46<y<-2.96$ (blue continuous line) and forward $2.03<y<3.53$ (red continuous line) rapidity intervals. .
}
\end{figure}

The ratio of both nuclear modification factors, i.e. $R_{pPb}^{\psi(2S)}/R_{pPb}^{J/\psi}$, also defined as the double ratio $\frac{[\psi(2S)/J/\psi]_{pPb}}{[\psi(2S)/J/\psi]_{pp}}$, is shown in Figure~\ref{fig:figratiopPbvncoll} for the backward and forward rapidity regions. In this double ratio the corrections due to the modification of the nuclear parton distribution functions cancel, and only comover interaction is at play.  
\vskip 0.15cm

We obtain a double ratio lower than one in both rapidity regions, due to the stronger effect of the comovers on the $\psi(2S)$. Moreover, this decrease below one is more pronounced in the backward rapidity region due to higher comover density which produces stronger dissociation on the $\psi(2S)$, due to its higher interaction cross section.
\vskip 0.15cm

\green{In the above results, 
the freeze-out time, $\tau_f$, is normalized by means of the pertinent density in p+p collisions at the corresponding energy. This leads
 in our approach to similar interaction times for RHIC d+Au and LHC p+Pb collisions.}
\green{It
harmonizes with the fact that the d+Au and p+Pb
systems, according to HBT measurements, are comparable in size \cite{Adare:2014vri,Adam:2015pya}.
}
\vskip 0.15cm

\green{However, as a limiting case and in order to take into account the possibility of a longer interaction time in the p+Pb case, we have also calculated the suppression if the p+Pb freeze-out density is decreased by a factor of 2, i.e. increasing the lifetime of a fireball by a factor of 2. We find that this could induce a stronger suppression, in particular for the excited states. This additional suppression will depend on the centrality of the collision and will be more important for the $\psi(2S)$ than for the $J/\psi$. In particular, while an additional 5\% over the original suppression value is obtained for the $J/\psi$,
the effect on the $\psi(2S)$ is of the order of 30\%.}
\vskip 0.15cm

In conclusion, we have performed a detailed 
study of $J/\psi$ and $\psi(2S)$ production in d+Au and p+Pb collisions at $\sqrt{s}=~200$~GeV and  $\sqrt{s}=5.02$ TeV respectively.
From our point of view, the available data constitute
experimental confirmation of 
the interaction of fully formed physical quarkonia 
with the produced particles --the comovers-- 
that happen to travel along with the $c\bar{c}$ pair.
\vskip 0.15cm

\red{
At the studied energies and rapidities, 
most of the physical bound states are formed well outside the target nucleus due
to the Lorentz dilation of the formation times  --in other words, the incoming nucleus is contracted by the Lorentz factor--.
This implies that 
the nucleons have swept over the nascent charmonium state, resulting in a nuclear absorption that can contribute little to the
charmonium suppression, and cannot account
for the difference between the
$J/\psi$ and the $\psi(2S)$ yields.}
\vskip 0.15cm
 
\red{On the other hand, the comovers can continue to interact well outside
of the nuclear volume, playing an important role.
The $J/\psi$ and $\psi(2S)$ produced outside the nucleus are surrounded by a dense system of hadrons (mainly pions) and converts into open charm due to interactions in the medium.}
In particular, the comover suppression can explain the relative modification of the $\psi(2S)$ to the $J/\psi$, 
$R_{pPb}^{\psi(2S)}/R_{pPb}^{J/\psi}$, in proton(deuteron)-nucleus collisions at RHIC and LHC energies. 
Other cold nuclear matter effects, as 
the nuclear modification of the parton distribution functions, cannot account for this difference \red{either}
since they impact similarly the $J/\psi$ and the $\psi(2S)$ .
\vskip 0.15cm

The comover effect, found to be of the order of 10\% in proton-nucleus collisions at SPS energy, increases with the  total particle multiplicity and achieves significant influence in proton(deuteron)-nucleus collisions at RHIC and LHC energies, in particular in the A-going direction.
\vskip 0.15cm

\green{
Our initial comover densities are proportional to the number of created hadrons.
In the transverse space, these densities, for minimum bias collisions, i.e. averaged over $b$, in the mid rapidity region are: $\rho^{co} \approx 4$ fm$^{-2}$  for p+Pb collisions at 5.02 TeV, and  $\rho^{co} \approx 2.5$ fm$^{-2}$ for d+Au collisions at 200 GeV.
These numbers agree with the multiplicities quoted above when they are divided by the transverse area over which comovers are produced, which corresponds to the collision overlap area, approximately equal to $\sigma^{inel}_{pp}=70$ mb for p+Pb collisions at 5.02 TeV,
and  $\sigma = \pi\ R_d^2$ for d+Au collisions at 200 GeV. In this later case,
due to the asymmetric nature of the deuteron in itself, this
overlap area can range between one to two times the p+p value at 200 GeV ($\sigma^{inel}_{pp}=40$ mb).
We take $R_d^2=3.2^{2/3}$ fm$^2$ \cite{Adare:2013nff,Klein-Bosing:2014uaa}
resulting in $\sigma = \pi\ R_d^2=68$ mb at 200 GeV. 
Note that our size parameters
are very similar to the experimental HBT radii \cite{Adare:2014vri,Adam:2015pya,Adare:2013nff,Abelev:2014pja} and to other theoretical estimates \cite{Klein-Bosing:2014uaa,McLerran:2013oju,Bzdak:2013zma}.}
\vskip 0.15cm

\green{
In order to put these quantities into context with equilibrium matter, it is interesting to study the 3-dimensional densities of comovers, which correspond simply to 
$\rho^{co}(\tau)=\rho^{co}/\tau$. 
Taking $\tau_0=1$ fm, equivalent to the 
formation time for the soft particles, one obtains 3-dimensional initial densities of the order of 4 fm$^{-3}$ in p+Pb LHC collisions and 2.5 fm$^{-3}$ in d+Au RHIC collisions.
These estimates can be compared with the estimate of the particle density in a
quark-gluon plasma \cite{Ollitrault:2007du}, 
$n=\frac{g}{\pi^2 \hbar^3} T^3$
where $g$ is the number of degrees of freedom (spin+colour+flavour), 16 for gluons and
24 for light u and d quarks, i.e. $g \approx 40$. Lattice QCD predicts that the transition to the quark-
gluon plasma occurs near $T_c \approx 190 - 170$ MeV \cite{Karsch:2007zza}.
Since $\hbar c = 197$ MeV fm, 
one obtains $n \approx 3.75 - 2.60$ fm$^{-3}$ at $T_c$.
One sees that the system is above the critical density only in p+Pb LHC collisions
if $\tau <1-1.5$ fm. In other words, the lifetime of an eventual quark-gluon plasma would be approximately 1 fm.}
\vskip 0.15cm

\green{
We are aware of the existence of hints of collective effects in p+Pb collisions at 5.02 TeV. In particular CMS \cite{Khachatryan:2010gv,CMS:2012qk} and ALICE  \cite{Abelev:2012ola} Collaborations
reported a near-side ridge, i.e. enhanced emission of particles with
similar azimuthal angles and different pseudorapidities developing in high-multiplicity p+p and p+Pb collisions, that can be associated with hydrodynamic flow.
 When a
fireball of strongly interacting matter is formed, azimuthal
correlations due to collective flow appear in the interaction. The interpretations are mostly based on collective harmonic flow \cite{Bozek:2012gr} 
or on the color-glass
condensate approach for the initial state \cite{Dusling:2012cg,Dusling:2012wy}.
Moreover, in \cite{Liu:2013via}, final-state hot medium effects, inspired on the formation of a quark-gluon plasma, have been proposed to study the quarkonium production 
 in p+Pb collisions at 5.02 TeV.
 }
 \vskip 0.15cm
 
 \green{
 Here we propose another mechanism, the final state interactions with comovers.
It should be stressed that 
the interaction of comovers at the early time $\tau_0$ involves large densities, 
which appear high for free mesons.
Hadronic matter in this situation is certainly far away from the ideal pion gas, but can be approximated as 
pre-hadrons or dressed mesons \cite{Linnyk:2008hp}, i. e. spectral densities with the quantum numbers of the hadronic states that show up at high
energy density, above 1 GeV fm$^{-3}$. As known from lattice QCD \cite{Karsch:2002wv,Karsch:2001uw,Petreczky:2003iz,Petreczky:2012rq},
the correlators for pions can
 survive above this energy density.
Such mesons are expected to have a more compact size in space. 
 }
 \vskip 0.15cm

We would like to thank Alfons Capella, Frederic Fleuret, Jean-Philippe
Lansberg, Michael Leitch and Ralf Rapp for stimulating and useful discussions and  Francois Arleo, Roberta Arnaldi, Zaida Conesa del Valle, Torstem Dahms, Anthony Frawley, Ramona Vogt, Peter Petreczky and Enrico Scomparin for insightful comments. We thank the Institute for Nuclear Theory at the University of Washington for its hospitality and partial support during the completion of this work. We also thank the Institut de Physique Nucl\'eaire de l'Universit\'e Paris-Sud and Laboratoire LePrince-Ringuet de l'Ecole Polytechnique for hospitality. This work was supported by the Ministerio de Economia y Competitividad of Spain.

\end{document}